\pdfoutput=1
\documentclass[english,manuscript]{aastex}
\usepackage[T1]{fontenc}
\usepackage[latin9]{inputenc}
\usepackage{graphicx}

\makeatletter

\shorttitle{Surface-shear shaped solar dynamo}
\shortauthors{Pipin \& Kosovichev}
\bibpunct[,]{(}{)}{;}{a}{,}{,}

\makeatother

\usepackage{babel}

\begin{document}

\title{The subsurface-shear shaped solar $\alpha\Omega$ dynamo}

\author{V.V. Pipin$^{1-3}$ and A.G. Kosovichev$^{3}$}

\affil{ $^{1}$ Institute of Geophysics and Planetary Physics, UCLA, Los
Angeles, CA 90065, USA \\
 $^{2}$Institute of Solar-Terrestrial Physics, Russian Academy
of Sciences, \\
 $^{3}$Hansen Experimental Physics Laboratory, Stanford University,
Stanford, CA 94305, USA }
\begin{abstract}
We propose a solar dynamo model distributed in the bulk of the convection
zone with  toroidal magnetic-field  flux   concentrated in
a near-surface layer. We show that if the boundary conditions at
the top of the dynamo region allow the large-scale toroidal magnetic
fields to penetrate close to the surface, then the modeled
butterfly diagram for the toroidal magnetic field in the upper
convection zone is formed by the sub-surface rotational shear
layer. The model is in agreement with observed properties of the
magnetic solar cycle.
\end{abstract}


\section{Introduction}

\label{sect:intro}

It is widely believed that the 11-year sunspot activity is produced
and organized by  large-scale magnetic fields generated somewhere
in the deep convection zone. Most of the solar
dynamo models suggest that the toroidal magnetic field that emerges on
the surface and forms sunspots is generated near 
the bottom of the convection zone, in the tachocline or just
beneath it in a convection overshoot layer, (see, e.g.,
\citealp{choud-shu-dik,bra-rue:05,dikchar,bonnano,tob-wei:07}). 
The belief in a  deep-seated solar dynamo comes from the fact that
this region is sufficiently stable, and can store magnetic flux despite the
magnetic flux-tube buoyancy effect
\citep{par:75,spie-weiss:80,vball:82,spruit-rob:83,vball-choud:88,choud:90}. 
The tachocline represents a strong radial shear of the angular velocity. Yet, 
turbulent diamagnetism (see, e.g., \citealp{zeld:57} or \citealp{kit:91})
pumps the magnetic fields from the intensively
mixed interior of convection zone to its boundaries. This effect can substantially
amplify the toroidal magnetic fields near the  convection zone
boundaries, (see, e.g.,
\citealp{kriv,guer:08}).

However, an attention was drawn to a number of theoretical and observational
problems concerning the deep-seated dynamo models \citep{bra:05,bra:06}.
A renewed discussion of  the place of
the solar dynamo can be found, e.g., in papers by \cite{bra:05} and \cite{tob-wei:07}.
In particular, there are some arguments that the sub-surface angular
velocity shear
could play an important role in the dynamo distributed
in the convection zone. This shear layer becomes an important ingredient
of the flux-transport dynamo models as well, (see, e.g., \citealp{guer:08}).

In this Letter we discuss the importance of the surface boundary conditions
for the dynamo models, which include the subsurface shear layer. 
The boundary conditions commonly used 
in the dynamo models correspond to a perfect conductor at the bottom of
 the convection zone and vacuum boundary conditions at the top.
Both the vacuum and perfect conductor boundary
conditions can be regarded as a mathematically convenient idealization.
The top boundary conditions  play a particularly important role
because they control the escape of the dynamo generated magnetic
fields from the Sun.

The perfect conductor boundary condition is usually identified as {}``closed'',
(e.g., \citealp{choud:84}), because in this case there is no penetration of the
generated magnetic flux to the outside. For the axi-symmetric
magnetic fields all magnetic field flux is closed inside the dynamo
region. The vacuum boundary condition is identified as {} ``open''.
In this case the poloidal field lines
are open to the outside, and the corresponding poloidal magnetic flux {}``freely''
escapes. Also, the strength of the toroidal magnetic field goes smoothly
to zero at the boundary. This means
that the vacuum boundary condition does not allow to 
the toroidal field penetrate to the surface. With such  boundary
condition it is hardly possible to form  sunspots from the
near-surface  large-scale toroidal magnetic fields.

Bearing in mind the dynamical nature of  magnetic fields on the solar surface
one can model the near-surface behaviour by using a combination
of the {}``open'' and {}``closed'' types of the boundary conditions.
Various consequences of this idea were explored, (see, e.g.,
\citealp{choud:84,reza:95,covas:98,kit-mazur:99,kit:00,reza:02,kap:10}).
Here, we apply this approach to a solar dynamo model that extends from the
bottom of the convection zone to the top, including the region of the strong sub-surface
rotational shear. We show that allowing  the toroidal magnetic flux to
penetrate
to the surface brings the butterfly diagram of the toroidal
large-scale magnetic field 
in the upper convection zone and also the phase relations between the different
components of the dynamo-generated magnetic field in agreement
with  solar-cycle observations.

\section{Dynamo equations}

\begin{figure*}
\begin{centering}
\includegraphics[scale=0.7]{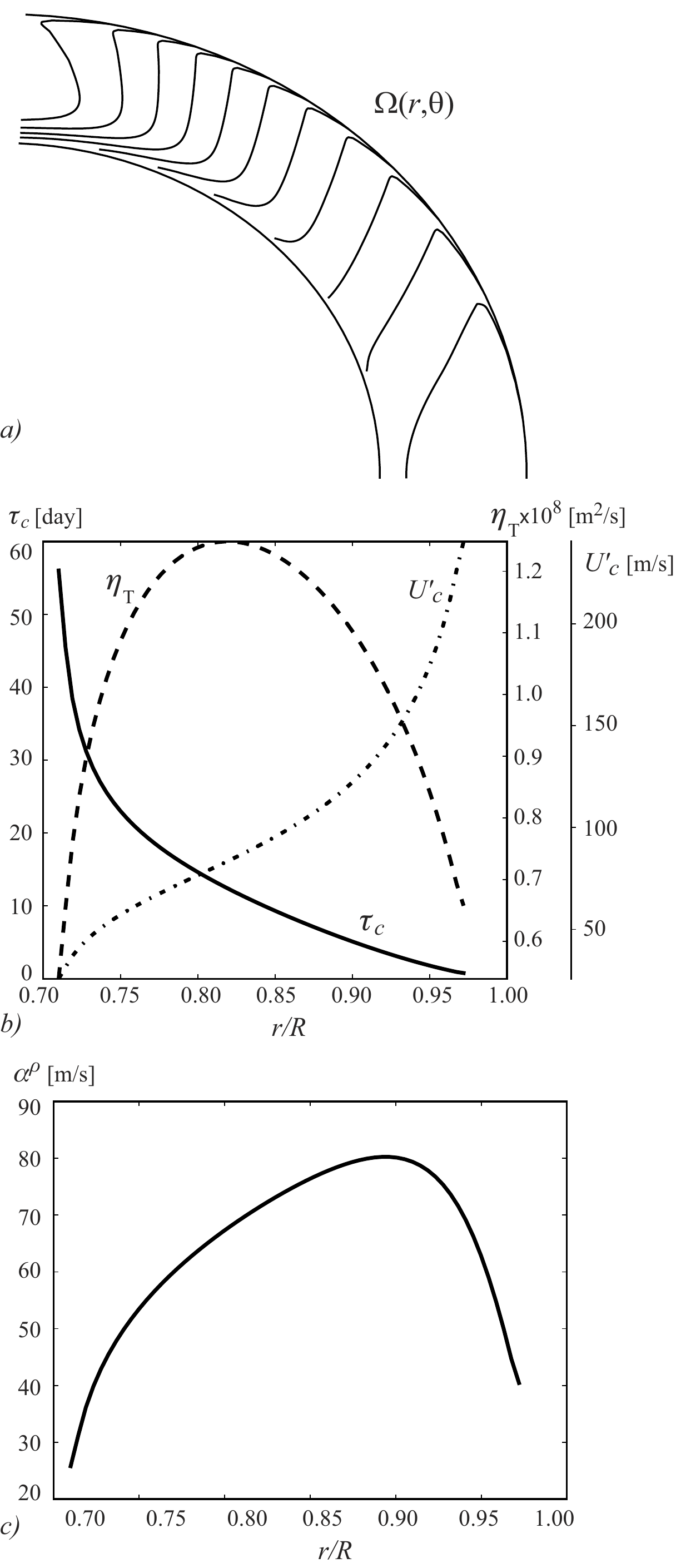}
\par\end{centering}

\caption{\label{fig:fig0-1-1} Internal parameters of the solar
  convection zone: 
a) { the  contours of the constant angular velocity are plotted for
  the levels $(0.75 - 1.05)\Omega_0$} with a step of $0.025\Omega_0$,
see, Eq.(\ref{eq:rotBA});
b) turnover convection time $\tau_c$, turbulent
  diffusivity $\eta_T$, RMS convective velocity $U'_c$; c) the
  radial profile of the dynamo $\alpha$-effect, 
$\alpha^{\rho}=\eta_T G f^{(a)}_{10}$, see
Eq.(\ref{eq:e_phi}).  The distance is measured in units of the solar radius.} 
\end{figure*}

The evolution of the axi-symmetric magnetic field ($B$ being the azimuthal
component of the magnetic field, $A$ is proportional to the azimuthal
component of the vector potential) is governed by the following equations:
\begin{eqnarray}
\frac{\partial A}{\partial t} & = &
r\sin\theta\mathcal{E}_{\phi}, \label{eq:A}\\
\frac{\partial B}{\partial t} & = &
-\sin\theta\left(\frac{\partial\Omega}{\partial r}\frac{\partial
    A}{\partial\mu}-\frac{\partial\Omega}{\partial\mu}\frac{\partial
    A}{\partial r}\right)+\frac{1}{r}\frac{\partial
  r\mathcal{E}_{\theta}}{\partial
  r}+\frac{\sin\theta}{r}\frac{\partial\mathcal{E}_{r}}{\partial\mu}\label{eq:B},
\end{eqnarray}
where,
\begin{eqnarray}
r\sin\theta\mathcal{E}_{\phi} & = & \psi_{\eta}\eta_{T}\left\{ f_{2}^{(d)}+2f_{1}^{(a)}\right\} \frac{\partial^{2}A}{\partial r^{2}}+\psi_{\eta}\eta_{T}\left\{ f_{2}^{(d)}+2f_{1}^{(a)}\right\} \frac{\left(1-\mu^{2}\right)}{r^{2}}\frac{\partial^{2}A}{\partial\mu^{2}}+\label{eq:e_phi}\\
 & + & \psi_{\alpha}C_{\alpha}\eta_{T}Gf_{10}^{(a)}r\mu\sin\theta f\left(\theta\right)B\nonumber \\
 & + & G\left(2f_{1}^{(a)}\mu\sin^{2}\theta\frac{\partial A}{\partial\mu}-\left(f_{3}^{(a)}-f_{1}^{(a)}\sin^{2}\theta\right)\frac{\partial A}{\partial r}\right)\nonumber \\
\mathcal{E}_{r} & = & \frac{\eta_{T}\psi_{\eta}}{r}\left\{ \left(f_{2}^{(d)}+2f_{1}^{(a)}\left(1-\mu^{2}\right)\right)\frac{\partial\sin\theta B}{\partial\mu}\right.\label{eq:e_r}\\
 & + & \left.2f_{1}^{(a)}\mu\sin\theta\frac{\partial}{\partial r}rB+2rGf_{1}^{(a)}\mu\sin\theta B\right\} \nonumber \\
r\mathcal{E}_{\theta} & = & \eta_{T}\psi_{\eta}\left\{ \left(f_{2}^{(d)}+2f_{1}^{(a)}\mu^{2}\right)\frac{\partial rB}{\partial r}\right.+\label{eq:e_th}\\
 & + & 2f_{1}^{(a)}\mu\sin\theta\frac{\partial}{\partial\mu}\left(\sin\theta B\right)\nonumber \\
 & - &
 \left.rG\left(f_{1}^{(a)}\cos2\theta+f_{3}^{(a)}\right)B\right\}
 \nonumber
 \end{eqnarray}
These equations are  similar to those used by 
\cite{pip-see09} and \cite{see-pip09}.
We use the same notations for the functions and parameters as in the
paper of \cite{pip08} (hereafter, P08). Here, $G=\partial_{r}\log\rho$
is the density stratification scale. Functions $f_{1,2,3,10}^{(a,d)}$ depend
on the Coriolis number $\Omega^{*}=2\tau_{c}\Omega_{0}$; functions
 $\psi_{\eta,\alpha}$ describe magnetic quenching
and depend on $\beta=B/\sqrt{\mu_{0}\rho\bar{u^{2}}}$. For
reference, these functions are given in Appendix. The parameter
$C_{\alpha}$ controls the strength of the  $\alpha$-effect. In the
 presented model the $\alpha$-effect is distributed in the bulk of the
 convection
zone. For a more clear
demonstration of the boundary condition impact, we confine the
$\alpha$-affect in a low-latitude region where the radial gradient of
the angular velocity is positive in the most part of the solar
convection zone. Similarly to \citet{dikp:04} we specify the
confinement function:
 \begin{equation}
 f(\theta)=\left(1+\mathrm{e}^{30\left(\left|\theta-\pi/2\right|-\pi/6\right)}\right)^{-1}.\label{conf-alph}
\end{equation}
In the  radial direction the $\alpha$-effect depends on the density
stratification, $G$, and function of the Coriolis number
$f^{(a)}_{10}\left(\Omega^*\right)$.
We introduce parameter
$C_{\eta}$ to control the turbulent diffusion coefficient, $\eta_{T}=C_{\eta}\eta_{T}^{(0)}$,
where $\eta_{T}^{(0)}=\tau_{c}\bar{u^{2}}/3$. The internal parameters
of the solar convection zone are given by \citet{stix}. At the top
 of the solar convection zone the stratification is strongly
deviates from adiabatic, and also the turbulence parameters vary sharply.
For this reason we confine the integration
domain between 0.71$R_{\odot}$ and 0.972$R_{\odot}$
in radius, and  it extends from the pole to pole in latitude. The differential
rotation profile, $\Omega=\Omega_{0}f_{\Omega}\left(x,\mu\right)$
(shown in Fig.\ref{fig:fig0-1-1}a)
is a slightly modified version of the analytical approximation  proposed by \citet{antia98}:
\begin{eqnarray}
f_{\Omega}\left(x,\mu\right) & = & \frac{1}{\Omega_{0}}
\left[\Omega_{0}+55\left(x-0.7\right)\phi\left(x,x_{0}\right)\phi\left(-x,-0.96\right)\right.\label{eq:rotBA}\\
 & - & \left.200\left(x-0.95\right)\phi\left(x,0.96\right)\right)\nonumber \\
 & + &
 \left(21P_{3}\left(\mu\right)+3P_{5}\left(\mu\right)\right]\left(\frac{\mu^{2}}{j_{p}\left(x\right)}
+\frac{1-\mu^{2}}{j_{e}\left(x\right)}\right)/\Omega_{0}\nonumber \\
j_{p} & = & \frac{1}{1+\exp\left(\frac{0.709-x}{0.02}\right)},\,\,
j_{e}=
\frac{1}{1+\exp\left(\frac{0.692-x}{0.01}\right)}\nonumber 
\end{eqnarray}
 where $\Omega_{0}=2.87\cdot 10^{-6}s^{-1}$ is the equatorial angular velocity of the Sun at the
surface, $x=r/R_{\odot}$, $\phi\left(x,x_{0}\right)=0.5\left[1+\tanh\left[100(x-x_{0})\right]\right]$,
$x_{0}=0.71$. The distribution of the Coriolis number, turbulent diffusivity
and the RMS convection velocity are shown in
Fig.\ref{fig:fig0-1-1}b. The radial profile of the $\alpha$-effect is
shown in Fig.\ref{fig:fig0-1-1}c.

At the bottom of the integration domain we apply the perfect
conductor ({} ``closed'')
boundary conditions: $\mathcal{E}_{\theta}=0,\, A=0$. The boundary
conditions at the top are defined as the following. Bearing in mind the
idea of  the partial escape of the toroidal flux from the Sun
discussed in Introduction,
 we explore a combination of  the {}``open'' and  {}``closed'' boundary conditions
at the top, controlled by a parameter $\delta$. 
For the toroidal field we use condition:
\begin{equation}
\delta\frac{\eta_{T}}{r_{e}}B+\left(1-\delta\right)\mathcal{E}_{\theta}
=  0 .\label{eq:tor-vac}
\end{equation}
This is similar to the boundary condition  discussed by \cite{kit:00}.
 For the poloidal field we apply
a combination of the local condition $A=0$ and condition of smooth
transition from the internal
poloidal field  to the external potential (vacuum) field: 
\begin{equation}
\delta\left(\left.\frac{\partial A}{\partial r}\right|_{r=r_{e}}
-\left.\frac{\partial A^{(vac)}}{\partial r}\right|_{r=r_{e}}\right)+\left(1-\delta\right)  A=  0,\label{eq:pol-vac}
\end{equation}
 where the external potential field is : 
\begin{equation}
A^{(vac)}\left(r,\mu\right)=
\sum
a_{n}\left(\frac{r_{e}}{r}\right)^{n}\sqrt{1-\mu^{2}}P_{n}^{1}\left(\mu\right),\label{eq:vac-dec}
\end{equation}
 $P_{n}^{1}\left(\mu\right)$ is the associated Legendre polynomial of
 degree $n$.
For the numerical implementation of Eq.(\ref{eq:pol-vac}), we
take a one-side finite difference approximation for the radial
derivative at an angular mesh point $\mu_j$:
\[\displaystyle \left.\frac{\partial A_{j}}{\partial
    r}\right|_{r=r_{e}}
=\frac{3A_{N\, j}-4A_{N-1\, j}+A_{N-2\, j}}{2h_{r}},
\]
where $h_r$ is the radial discretization interval, and consider the
expansion
 (\ref{eq:vac-dec}) at the top boundary $r=r_{e}$:
$A_{N\, j}=\sum
a_{n}\sqrt{1-\mu_{j}^{2}}P_{n}^{1}\left(\mu_{j}\right)$.
Then, we define matrices 
$M_{nj}^{\left(a\right)}=\sqrt{1-\mu_{j}^{2}}P_{n}^{1}\left(\mu_{j}\right)$
and
$\tilde{M}_{jn}=n\sqrt{1-\mu_{j}^{2}}P_{n}^{1}\left(\mu_{j}\right)$,
with $\mu_j$ being collocation points of $P_{n}^{1}$ (see, \citealp{boyd}). 
This procedure allows us to express  coefficients $a_n$ in (\ref{eq:vac-dec})
via the values of potential $A$ at the grid-points: $a_{n}=M_{nj}^{\left(a\right)-1}A_{N\, j}$.
Substituting this in Eq.(\ref{eq:pol-vac}) and solving it we get : 
\[
A_{N\, j}=\delta\left(3E_{jk}+2h_{r}\tilde{M}_{jk}M_{nk}^{\left(a\right)-1}\right)^{-1}\left(4A_{N-1\, jk}-A_{N-2\, k}\right),\]
 where $E$ is a unit diagonal matrix. 

\section{Results and discussion}

Parameter $\delta$ in the boundary conditions describes a transition
between the ``closed'' ($\delta=0$) to ``open'' ($\delta=1$)
boundaries. Physically, it controls penetration of the dynamo-generated
fields into the outer atmosphere.

Decreasing $\delta$ in Eqs.(\ref{eq:tor-vac},\ref{eq:pol-vac}) results
in stronger tangential and  weaker radial large-scale magnetic
fields at the surface.
While the strong toroidal magnetic field is a desired feature of 
the model, the weak radial magnetic field decreases the efficiency
of the radial subsurface shear to produce large-scale toroidal magnetic fields.
In fact, the simulations reveal that the critical dynamo number, $C_{\alpha}$,
is greater when the penetration parameter $\delta$ { is smaller}. For this
reason we consider the case of a small deviation 
 from the vacuum ({} ``open'') boundary conditions.
 Moreover, in order to match the dynamo
period to the solar cycle  we choose the magnetic diffusivity
parameter $C_{\eta}=0.05$,
which is significantly lower then the value predicted by the mixing length
theory. To demonstrate the
effect of the new boundary conditions with the field penetration we
show for comparison in Figure 2 and 3 the results
of two runs for  $\delta=0.95$
(corresponding to a partial penetration of toroidal field) and 
$\delta=1$ (the vacuum boundary conditions). 
\begin{figure*}
\begin{centering}
\includegraphics[width=0.7\textwidth]{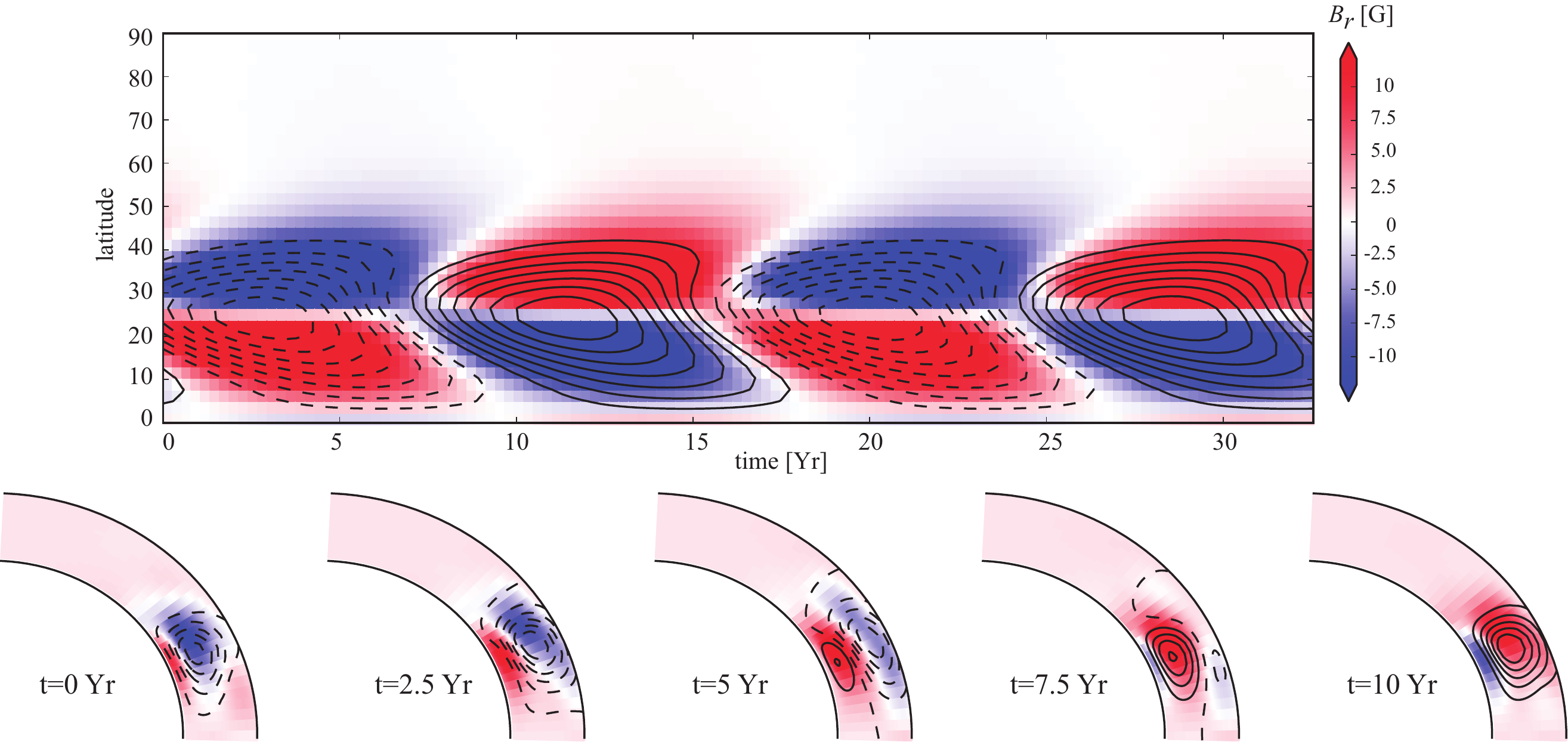} 
\par\end{centering}

\caption{\label{caseI}The case of $\delta=0.95$ (the top boundary conditions with
a partial penetration of toroidal magnetic fields into the outer layers of
the Sun). The top panel shows the near-surface
toroidal component of large-scale magnetic fields (contour lines) and
the surface radial component of the field (color background).
The bottom panel shows snapshots  of the poloidal (contour
lines) and toroidal magnetic field  components for a half of the magnetic cycle.
 The maximum strength of the toroidal field is about 1KG. Time is in years.}

\end{figure*}

\begin{figure*}
\begin{centering}
\includegraphics[width=0.7\textwidth]{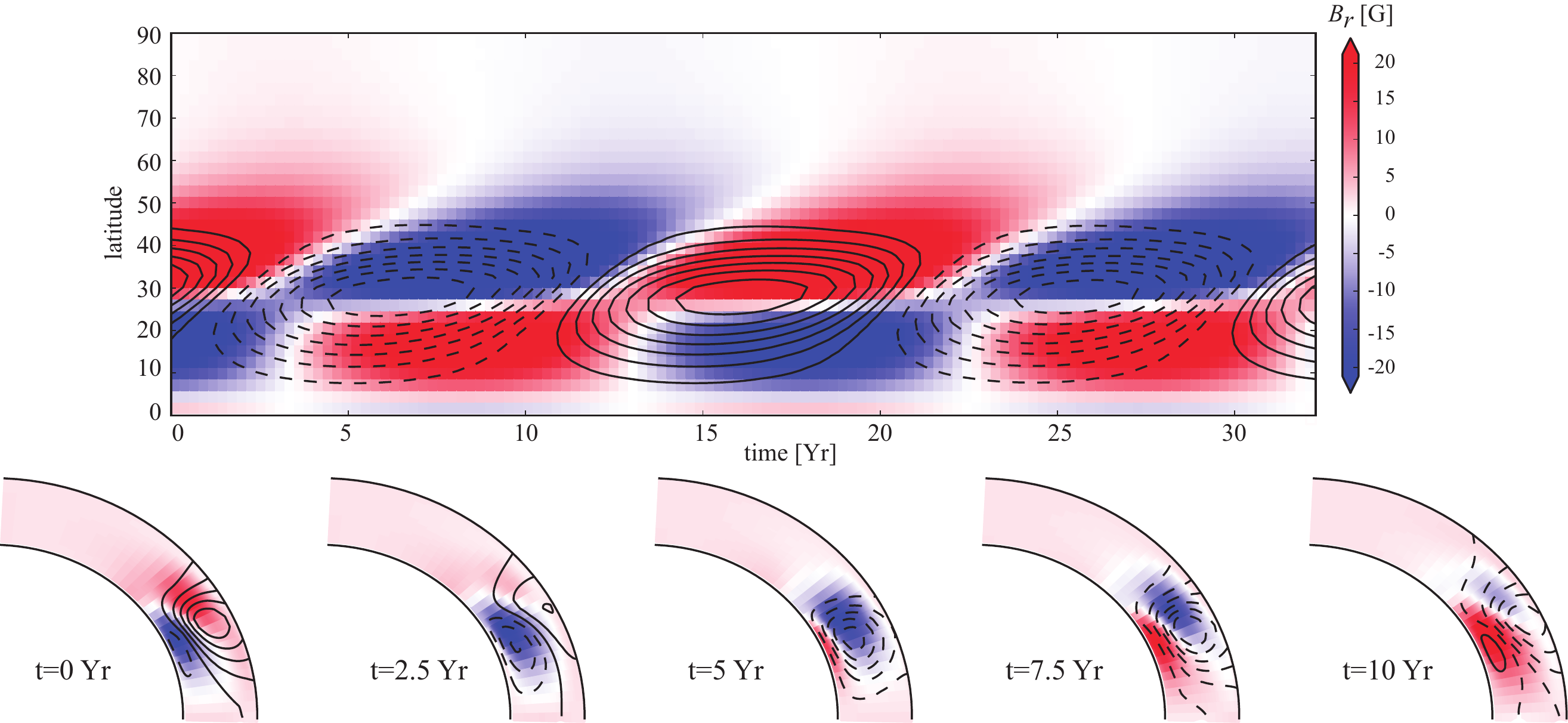} 
\par\end{centering}

\caption{\label{caseII} The same as in on the Fig.\ref{caseI} for the vacuum
boundary conditions, $\delta=1$.}
\end{figure*}

These results show that allowing the large-scale toroidal
magnetic field to penetrate in to the surface layers of the Sun changes the
direction of the latitudinal migration of the toroidal field activity
and produces the magnetic butterfly diagram  in a good qualitative agreement with
 solar-cycle observations. The dynamo-wave penetrates to the
surface and propagates along the iso-surface of angular velocity in
the subsurface shear layer. This is
in agreement with the Yoshimura rule \citep{yosh}.

{ \cite{dikp:02} explored generation of toroidal magnetic fields 
by the $\Omega$-effect in the sub-surface shearlayer in the Babcok-Leighton-type
  dynamo models. They found that  the phase relation between the sub-surface
  toroidal magnetic field and 
the surface  radial magnetic field is inconsistent with observations.
 We believe that this inconsistency was  because in their model the
 source of the surface poloidal magnetic fields was related with 
 the bottom of convection zone. Therefore, the sub-surface toroidal
 field that is generated in the  sub-surface shear layer does not
 contribute directly to the generation of the poloidal magnetic field.}
 
Both of our simulation runs were started
with the initial magnetic fields with of the equally mixed symmetrical
and antisymmetrical realtive to the equator components.
 The evolution retains only the
dipole-like parity configurations in both cases though
the relaxation time in the penetration case of $\delta=0.95$ is much
longer than in the case of
the pure vacuum boundary conditions. If we relax the confinement of
the alpha-effect in latitude, i.e., $f(\theta)=1$ instead of  Eq.(\ref{conf-alph}),
the general patterns of Fig.\ref{caseI} are hold except that the
maximum of the toroidal magnetic field is shifted to higher latitude
$\approx 40^{\circ}$.
Therefore we can conclude that the $\alpha\Omega$- dynamo model with
the boundary
 conditions that allow a small partial penetration of the toroidal
 field into the outer layers of the Sun, can robustly reproduce
the solar-cycle butterfly diagram for the near-surface large-scale
magnetic field evolution. These results demonstrate the importance  of
the subsurface rotational shear layer in the solar dynamo mechanism.

\section{Acknowledgements}

This work was supported by the NASA LWS NNX09AJ85G grant and partially
by the RFBR grant 10-02-00148-a.


\begin{thebibliography}{}
\bibitem[{Antia} {et~al.}(1998)]{antia98} {Antia}, H. M. , {Basu},
Sarbani, {Chitre}, S. M. 1998, MNRAS, 298,543

\bibitem[{Boyd} (2001)]{boyd}
Boyd, J.P., 2001, Chebyshev and Fourie Spectral Methods, 2nd
ed.(Mineola, N.Y.: Dover Publications)

\bibitem[{Brandenburg}(2006)]{bra:06} Brandenburg, A. 2006, ASP Conference
Series, 354,Solar MHD Theory and Observations: A High Spatial Resolution
Perspective, 121

\bibitem[{Brandenburg}(2005)]{bra:05} Brandenburg, A.,2005, ApJ,
625, 539

\bibitem[{Brandenburg \& Subramanian}(2005)]{bra-sub:05} Brandenburg,
A., Subramanian, K..2005, Phys. Rep. 417, 1

\bibitem[{Bonanno} {et~al.}(2002)]{bonnano} Bonanno, A., Elstner,
D., R\"udiger, G., 2002, A\&A, 390, 673

\bibitem[{Choudhuri} {et~al.}(1995)]{choud-shu-dik} Choudhuri, A.
R., Sch\"ussler, M., \& Dikpati, M. 1995, A\&A, 303, L29

\bibitem[{Choudhuri}(1984)]{choud:84} Choudhuri, A. R. 1984, ApJ,
281, 846

\bibitem[{Choudhuri}(1990)]{choud:90} Choudhuri, A. R. 1990, ApJ,
355, 733

\bibitem[{Covas}{et~al.}(1998)]{covas:98} Covas, E., Tavakol, R.,
Tworkowski, A., Brandenburg, A. 1998, A\&A, 329, 350

\bibitem[{Dikpati \& Charbonneau}(1999)]{dikchar} Dikpati, M., \&
Charbonneau, P. 1999, ApJ, 518, 508

\bibitem[{Dikpati} {et~al.}(2002)]{dikp:02} Dikpati, M.,
Corbard, T., Thompson, M.J., Gilman, P. 2002, ApJ,575,L41

\bibitem[{Dikpati} {et~al.}(2004)]{dikp:04} Dikpati, M., de Toma,
G., Gilman, P.A., Arge, C.N., White, O.R. 2004, ApJ,601,1136


\bibitem[{Guerrero \& de Gouveia Dal Pino}(2008)]{guer:08} Guerrero,
G., de Gouveia Dal Pino, E. M. 2008, A\&A, 485, 267

\bibitem[{K\"apyl\"a} {et~al.} (2010)]{kap:10} 
K\"apyl\"a, P.J., Korpi,
M.J. \& Brandenburg, A. 2010, A\&A, 518, 22

\bibitem[{Kitchatinov \& R\"udiger}(1992)]{kit:91}
Kitchatinov, L.L., R\"udiger, G. 1992, A\&A,260, 494

\bibitem[{Kitchatinov \& Mazur}(1999)]{kit-mazur:99}
Kitchatinov, L.L., Mazur, M.V. 1999, AstL,25, 471

\bibitem[{Kitchatinov }{et~al.}(2000)]{kit:00} Kitchatinov, L. L.,
Mazur, M. V., Jardine, M., 2000, A\&A, 359, 531



\bibitem[{Krivodubskij}(1987)]{kriv} Krivodubskij, V. N., 1987,
SvAL,13, 338


\bibitem[{Parker}(1975)]{par:75} Parker, E.N., 1975, ApJ, 198, 205

\bibitem[{Pipin \& Seehafer}(2009)]{pip-see09} {Pipin}, V.V. \&
{Seehafer}, N. 2009, A\&A, 493,819

\bibitem[{Pipin} (2008)]{pip08}Pipin, V. V. 2008, Geophys. Astrophys.
Fluid Dynam., 102, 21

\bibitem[{R\"udiger \& Brandenburg}(1995)]{bra-rue:05}
 R\"udiger, G., Brandenburg, A., 1995, A\&A, 296, 557

\bibitem[{Seehafer \& Pipin}(2009)]{see-pip09} {Seehafer}, N.
\& {Pipin}, V.V. 2009, A\&A, 508,9

\bibitem[{Spiegel \& Weiss}(1980)]{spie-weiss:80} Spiegel, E.A.,
\& Weiss, N.O., 1980, Nature, 287, 616

\bibitem[{Spruit \& Roberts}(1983)]{spruit-rob:83} Spruit, H.C.,
\& Roberts, B., 1983, Nature, 304, 401

\bibitem[{Stix}(2002)]{stix} Stix, M. 2002, The Sun. An Introduction,
2nd Ed. (Berlin: Springer)

\bibitem[{Tavakol }{et~al.}(1995)]{reza:95} Tavakol, R.K., Tworkowski,
A. S., Brandenburg, A., Moss, D., Tuominen, I., 1995, A\&A, 296,
269

\bibitem[{Tavakol }{et~al.}(2002)]{reza:02} Tavakol, R.; Covas,
E.; Moss, D.; Tworkowski, A. 2002 A\&A, 387, 1100

\bibitem[{Tobias \& Weiss}(2007)]{tob-wei:07} Tobias, S. \& Weiss,
N. 2007, in {}``The Solar Tachocline'', eds by Hughes,D.W., Rosner,
R. and Weiss N.O., CUP, Cambridge, UK, p.319

\bibitem[{van Ballegooijen}(1982)]{vball:82} van Ballegooijen, A.
A. 1982, A \& A, 113, 99

\bibitem[{van Ballegooijen \& Choudhuri}(1988)]{vball-choud:88}
van Ballegooijen, A.A., Choudhuri, A.R. 1988, ApJ, 333, 965

\bibitem[{Zeldovich}(1957)]{zeld:57}
Zeldovich, Ya.B. 1957, Sov.Phys. JETP, 4, 460

\bibitem[{Yoshimura}(1975)]{yosh} Yoshimura, H. 1975 ApJ, 201, 740 

\end{thebibliography}

\subsection{Appendix}

Here, we give the definitions of the functions which were used in
the model. The given functions describe the efficiency of the Coriolis
force and the mean magnetic field to act on the stratified turbulence
and to produce the dynamo $\alpha$-effect, anisotropy of magnetic diffusion,
 turbulent magnetic pumping, magnetic quenching of the turbulent effects,
etc. These effects are discussed in (Pipin, 2008).  
Functions $f_{1,2,3,10}^{(a,d)}$ depend
on the Coriolis number $\Omega^{*}=2\tau_{c}\Omega_{0}$; functions
 $\psi_{\eta,\alpha}$ describe magnetic quenching
and depend on $\beta=B/\sqrt{\mu_{0}\rho\bar{u^{2}}}$:
\begin{eqnarray*}
f_{1}^{(a)} & = & \frac{3}{4\Omega^{*\,2}}\left(\left(\Omega^{*\,2}+3\right)\frac{\arctan\Omega^{*}}{\Omega^{*}}-3\right),\\
f_{2}^{(d)} & = & \frac{3}{4\Omega^{*\,2}}\left(\left(\left(\varepsilon-1\right)\Omega^{*\,2}+3\varepsilon+1\right)\frac{\arctan\left(\Omega^{*}\right)}{\Omega^{*}}-\left(3\varepsilon+1\right)\right),\\
f_{3}^{(a)} & = & \frac{3}{4\Omega^{*\,2}}\left(\left(\left(\varepsilon-1\right)\Omega^{*\,2}+\varepsilon-3\right)\frac{\arctan\Omega^{*}}{\Omega^{*}}+3-\varepsilon\right),\\
f_{10}^{(a)} & = & -\frac{1}{3\Omega^{*\,3}}\left(3\left(\Omega^{*2}+1\right)\left(\Omega^{*2}+\varepsilon-1\right)\frac{\arctan\Omega^{*}}{\Omega^{*}}-\left(\left(2\varepsilon+1\right)\Omega^{*2}+3\varepsilon-3\right)\right),\\
\psi_{\eta} & = & \frac{3}{16\beta^{2}}\left(\left(\left(4\beta^{2}+3\right)\varepsilon+4\beta^{2}-1\right)\frac{\arctan\left(2\beta\right)}{2\beta}+1-3\varepsilon\right)\\
\psi_{\alpha} & = &
\frac{5}{128\beta^{4}}\left(16\beta^{2}-3-3\left(4\beta^{2}-1\right)\frac{\arctan\left(2\beta\right)}{2\beta}\right).
\end{eqnarray*}
 The parameter
$\varepsilon$ measures the ratio between the turbulent energies of the kinetic and
magnetic fluctuations $\varepsilon=\bar{b^{2}}/ \mu_{0}\rho\bar{u^{2}}$, in the background turbulence (in
the absence of the mean-fields). Note, in notation of Pipin(2008), 
the turbulent diffusivity and $\alpha$-effect quenchning functions are defined as follows,
$\psi_{\eta}=\phi_{3}+\phi_{2}-2\phi_{1}$, and
$\psi_{\alpha}=-3/4\phi^{(a)}_6$, respectively. Expressions
for $\phi_{1,2,3}$ and $\phi^{(a)}_6$ are given in  (Pipin, 2008).
\end{document}